\documentstyle[12pt]{article}
\begin{document}
\baselineskip 22pt plus 2pt
\newcommand{\alt}{\raisebox{-0.5ex}{$\stackrel{<}\sim$}}
\newcommand{\limes}{ \parbox{.8cm}
{{\tiny $B_1 \rightarrow \nolinebreak B \vspace{-.2cm}\\ 
B_2 \rightarrow \nolinebreak  B$}}}
\rule{7cm}{0.0cm}
\begin{minipage}[b]{9cm}
\begin{small}
To appear in {\em Waves in Random Media} 
-  a topical issue on 
"Disordered Electronic Systems"
\end{small}
\end{minipage}\\

 \begin{center}
{\bf Orbital Magnetism in Disordered Mesoscopic Metals} \\ \ \\
B. Shapiro\\
Department of Physics, Technion, Haifa 32000 \\ \ \\
{\bf Abstract}\\ \ \\

\end{center}

The theory of orbital magnetism in disordered metals is reviewed, and 
extended to include a broad range of temperatures and fields. Sample-to-sample
fluctuations in the orbital magnetic susceptibility are studied. In a given 
sample
these fluctuations manifest themselves in aperiodic sample-specific 
oscillations of susceptibility and magnetization, when the strength of the
magnetic field is changed.

\noindent{\bf 1. Introduction}

A degenerate electron gas in a macroscopic metal exhibits weak orbital
magnetism (the Landau diamagnetism) [1]. There are two factors that can 
significantly enhance the orbital magnetic response: finite system size
and electron-electron interactions.  This paper is devoted primarily to 
reviewing and extending recent work on the finite size (i.e. mesoscopic)
effects on orbital magnetism of disordered metals (some of the results obtained
prior to 1993 have been reviewed in [2]).  The effect of interactions will
be briefly addressed at the end of the paper.

In a weakly disordered metal the elastic mean free path $\ell$
is much larger than the inverse Fermi number $k^{-1}_F$ but much smaller
than the sample size $L$. Under these conditions electrons propagate by a 
diffusion process, with a diffusion coefficient $D=v_F\ell/d$, where
$v_F$ is the Fermi velocity and $d$ is the sample dimensionality.
In addition to the three length scales $(k^{-1}_F$, $\ell$, $L$) which
characterize the sample, there are two more lengths, 
$L_T = (\hbar D/T)^{\frac{1}{2}}$ and 
$L_B = (\hbar c/eB)^{\frac{1}{2}}$, which specify the temperature $T$ and the
external magnetic field $B$. We assume that the inelastic scattering (or
the phase breaking) length is larger than $L_T$ and, thus, irrelevant for
our purpose.

Whether a sample of a given size $L$ can be considered
as macroscopic, depends on the temperature. If $T$ is smaller than the
Thouless energy 
$E_c\simeq \hbar D/L^2$ (i.e. $L_T > L)$, the sample forms a single
coherent unit and can exhibit large mesoscopic effects. The orbital
magnetic susceptibility $\chi$ fluctuates from sample to sample and the
typical fluctuation, $\langle \Delta \chi^2\rangle^{\frac{1}{2}}$, can
exceed the average value $\langle \chi\rangle$ by a large factor (angular
brackets denote averaging over an ensemble of macroscopically identical
samples). Mesoscopic effects in orbital magnetism are not restricted to
$T < E_c$ but persist to much higher temperatures, although their magnitude
gradually decreases [3-6]. A detailed study of \
$\langle \Delta \chi^2\rangle^{\frac{1}{2}}$,
in the temperature regime
$E_c \ll T \ll \hbar v_F/\ell$, was made in [5,6].
Some of these results will be re-derived below. The present treatment, however,
allows us to consider a much broader range of temperatures and fields.
In particular, temperatures larger than $\hbar v_F/\ell$ (i.e. $L_T<\ell$) 
and fields stronger than $\hbar c\ell^2/e$ (i.e. $L_B<\ell)$ will be
discussed.

In Sec.\ 2 we explain the phenomenon of mesoscopic orbital magnetism 
using an example of a clean chaotic cavity. A clear understanding of this
simple system is helpful for the qualitative considerations of Sec.~3.
A quantitative theory for
$\langle \Delta \chi^2\rangle^{\frac{1}{2}}$ of a disordered metal is
set up in Sec.~4, and its consequences are examined in Sec.~5,
for various temperatures and fields. Sec.~6 is devoted to a
brief discussion of the interaction effects. Conclusions are 
summarized in Sec.~7.\\

\noindent{\bf 2. Chaotic cavities}

Before turning to disordered metals, let us briefly discuss the simpler case of a 
clean chaotic cavity. We follow below the presentation in [2]. (Similar considerations
for persistent currents appear in [7,8].) The two relevant energies are the
level spacing $\Delta$ and the ``inverse time of flight'' across the cavity,
$\Gamma \simeq \hbar v_F/L$. The mesoscopic regime corresponds to the 
temperature range $\Delta \ll T \alt \Gamma$. In this regime the thermodynamics \
of the system is determined by the density of states
$\rho_{\Gamma} (\epsilon)$,
smoothed over an interval of order $\Gamma$ near energy $\epsilon$. The
main point is that $\rho_\Gamma$ contains a small oscillatory component,
$\delta \rho_\Gamma$, which is due to the shortest periodic orbit [9]
\begin{eqnarray}
\delta \rho_\Gamma (\epsilon, B)\simeq \frac{1}{\Gamma} \cos
(k L + \gamma) \cos \left(\frac{2\pi B L^2}{\phi_0}\right) \ . 
\end{eqnarray}
where $k=\hbar^{-1}\sqrt{2 m \epsilon}$, 
$\phi_0 =2\pi\hbar c/e$, and the magnetic  field $B$ is assumed to be weak,
so that the cyclotron radius $R_c=v_F mc /eB$ is larger than the size
$L$ of the cavity. The oscillatory term in  the density of states 
contributes
a small oscillatory correction $\delta \Omega$ to the thermodynamic potential:
\begin{eqnarray}
\delta\Omega = - T\int d\epsilon \delta\rho_\Gamma (\epsilon)
\ln \left[ 1+\exp \left(\frac{\mu-\epsilon}{T}\right)\right] \ , 
\end{eqnarray}
which, after integrating by parts twice and keeping the leading term
in the large parameter $k_FL$, gives 
\begin{eqnarray}
\delta \Omega \simeq \Gamma \cos (k_F L + \gamma) \cos \left(
\frac{2\pi B L^2}{\phi_0}\right) \ . 
\end{eqnarray}
It is this correction that, due to its sensitivity to the magnetic field, 
dominates the differential magnetic susceptibility
\begin{eqnarray}
\chi = - L^{-d} \frac{\partial^2\Omega}{\partial B^2} \simeq \chi_0
(k_FL)^{3-d} \cos(k_FL+\gamma)\cos \left(\frac{2\pi B L^2}{\phi_0}\right), 
\end{eqnarray}
where $\chi_0 \simeq e^2 k_F^{d-2}/mc^2$ is the absolute value of the
Landau susceptibility, in $d$ dimensions.

Thus, the orbital susceptibility of a two-dimensional electron gas
confined to a chaotic cavity (a quantum dot) is enhanced by a factor
$k_FL$, as compared to $\chi_0$. Detailed semiclassical computations of the
orbital magnetic susceptibility for a generic chaotic system have been
performed by Agam [10], and for a ``stadium billiard'' by Raveh [6]. Similar 
semiclassical considerations apply also to integrable systems [11-13] where,
due to existence of families of periodic orbits, the  enhancement is 
even larger. In all clean systems mesoscopic effects rapidly disappear when
temperature is  raised above $\Gamma$, i.e. for $T\gg\Gamma$ the system
becomes macroscopic and its orbital susceptibility assumes the Landau value.
This is in contrast to the situation in disordered metals, where the effect 
decays with temperature rather gradually.\\

\noindent{\bf 3. Disordered Metals: Qualitative Considerations}

The following qualitative discussion is confined to two-dimensional samples.
In a disordered metal electrons propagate diffusively, so that the energy
$\Gamma$ of the previous section should be replaced by the Thouless
energy $E_c=\hbar D/L^2$. For $T\alt E_c$, a change
$\delta B \simeq \phi_0/L^2$
in a magnetic field produces a change $\delta\Omega\simeq E_c$ in the 
thermodynamic potential. The typical magnetic susceptibility
$\chi_{\rm typ}$
of a given sample can have either sign, and its magnitude is of
order $\delta\Omega/L^2 \delta B^2\simeq \chi_0 k_F \ell$.

The large sample-to-sample fluctuations cancel out, to a large extent, if 
one averages over many samples, with different impurity arrangements. The
resulting average value $\langle \chi \rangle$ is much smaller than
$\chi_{\rm typ}$
and depends on whether the samples are in contact with particle 
reservoir (grand canonical system) or are isolated (a canonical one).
In the former case $\langle \chi \rangle \approx -\chi_0 = 
- e^2/12\pi mc^2$, whereas in  the latter there is an additional
paramagnetic contribution which, for $T\ll E_c$, is the dominant one
[4,14]. We will concentrate on the typical susceptibility, or, more 
precisely,
on its fluctuating part $\langle \Delta \chi^2\rangle^{\frac{1}{2}}
\equiv \delta \chi$. This quantity is insensitive to the type of 
thermodynamic ensemble.

Thus, for $T\alt E_c$, $\delta \chi \simeq \chi_0 k_F\ell$. Let us see
how $\delta \chi$ changes when $T$ is raised above $E_c$. For
$E_c\ll T \ll (\hbar v_F/\ell) \equiv \hbar/\tau$
(i.e. $\ell \ll L_T\ll L)$,
one can view the sample as made up of $(L/L_T)^2$ boxes, of size $L_T$
each. The above estimate is applicable for each box separately, i.e.
$\delta \chi_{\rm Box} \simeq \chi_0 k_F\ell$ and the 
fluctuating part of the magnetic moment of a box is $\delta M_{\rm Box}
\simeq \chi_0 k_F\ell B L^2_T$. Summing over all boxes, with random signs,
one obtains the typical fluctuation of the magnetic moment of the entire
sample, $\delta M \simeq \chi_0 k_F\ell B L^2_T(L/L_T)$, i.e.
$\delta \chi/\chi_0\simeq k_F\ell(L_T/L) \sim T^{-\frac{1}{2}}$. This
ratio is smaller than $k_F\ell$ but can be larger than unity. Even if this
relative fluctuation is small, it can still be of interest because of its
non-trivial temperature and field dependence (strong field effects 
will be discussed later).

The picture changes, if the temperature is raised further and becomes
larger than $\hbar/\tau$ (i.e. $L_T< \ell$).
The relevant trajectories, which dominate the magnetic response, are now of 
ballistic (rather than diffusive) nature, i.e. shorter than $\ell$. Only
short trajectories, of an area $S\alt (\hbar v_F/T)^2\equiv \ell^2_T$, can 
efficiently respond to the magnetic flux. Let us denote by 
$W_T$ the probability for having this kind of a trajectory, in a ``box''
of size $\ell_T$. The probability $P(S)dS$ for a closed ballistic trajectory, of
an area between $S$ and $S+dS$, was estimated in [15] as 
$P(S)dS \simeq \ell^{-3} \sqrt{S} dS$, so that $W_T\simeq (\ell_T/\ell)^3$.
The fluctuating part of the susceptibility $\delta \chi$ of the sample
is estimated in a way similar to that used above for the diffusive regime:
The sample is partitioned into $(L/\ell_T)^2$ boxes; a box contains,
with probability $W_T$, a ballistic trajectory of an area of order
$\ell_T^2$ which contributes a magnetic moment
$|\delta M_{\rm Box}|\simeq \chi_0 k_F\ell_T \ B\ell^2_T$
(the factor $k_F\ell_T$ is the enhancement factor discussed in Sec.~2).
Adding up the (random) contributions of all boxes, and remembering that
only a fraction $W_T$ of all boxes does carry magnetic moments, one
obtains $\delta M \simeq |\delta M_{\rm Box}|(W_TL^2/\ell^2_T)^{\frac{1}{2}}$,
i.e. $\delta \chi \simeq \chi_0 k_F\ell (\ell/L)(\ell_T/\ell)^{\frac{7}{2}}$.
Thus, the $T^{-\frac{1}{2}}$ dependence of the diffusion regime $(E_c < T < \hbar/\tau)$
crosses over to a $T^{-\frac{7}{2}}$ dependence in the ballistic regime
$(T > \hbar/\tau)$.

The arguments so far were restricted to weak magnetic fields, i.e.
to the regime of linear response. The region of validity of linear response
depends, of course, on temperature. For instance, in the diffusion regime
linear response requires $L_B>L_T$, i.e. $\hbar \omega_c < T/k_F\ell$, where
$\omega_c$ is the cyclotron frequency. For stronger fields various non-linear
effects in susceptibility and magnetization show up. In particular, for
$L_B < L_T$, one may expect aperiodic, sample-specific fluctuations in
magnetization [6]: every time when the magnetic field acquires an increment
$\Delta B \simeq \phi_0/L^2_T$, an effectively ``new sample'' is created
and, thus, the differential susceptibility changes by an amount
of order $\pm \chi_0 k_F\ell(L_T/L)$. These sample-specific fluctuations
resemble the universal conductance fluctuations (see [16] for an extensive
discussion), although the
relative fluctuation in susceptibility is much larger than in the 
conductance.\\

\noindent{\bf 4. Disordered Metals: Theory}

The grand potential, for a specific sample, is 
\begin{eqnarray}
\Omega = V_0 \int d \epsilon \nu (\epsilon, B) F_T(\epsilon-\mu), 
\end{eqnarray}
where $V_0 = L^d$ is the sample volume, $\nu(\epsilon, B)$ is  the density
of states at field $B$, per unit volume, and
\begin{eqnarray}
F_T(\epsilon-\mu)=- T\ln\left[1+\exp\left(-\frac{\epsilon-\mu}{T}\right)
\right]. 
\end{eqnarray}
The sample magnetization is $M=-V_0^{-1}(\partial \Omega/\partial B)_{\mu,V,T}$
and its differential susceptibility is $\chi = \partial M/\partial B$. The
variance, $\langle \Delta \chi^2\rangle$, in the ensemble of random samples
is therefore
\begin{eqnarray}
\langle \Delta \chi^2\rangle
= \lim_{B_2\rightarrow B}\lim_{B_1\rightarrow B}
\frac{\partial^4}{\partial B_1^2\partial B^2_2}
\int\!\!\int d\epsilon_1 d\epsilon_2 F_T
(\epsilon_1-\mu) F_T(\epsilon_2-\mu)
K(\epsilon_1,\epsilon_2;B_1,B_2), 
\end{eqnarray}
where $K(\epsilon_1,\epsilon_2;B_1,B_2)=\langle \Delta\nu(\epsilon_1,
B_1)\Delta\nu(\epsilon_2,B_2)\rangle$ is the correlation function for
the density of states, at two different energies and for two close
values of the magnetic field.

The density of states correlation function for diffusive metals was
first considered by Altshuler and Shklovskii [17]. In the context of
orbital magnetism it has been studied by a number of authors [4-6,14,18].
The computations are based on writing the density of states in terms of
retarded and advanced Green's functions:
\begin{eqnarray}
\nu(\epsilon, B) = - s \frac{i}{2\pi V} 
Tr\left[ G_R(\epsilon,B)-G_A(\epsilon,B)\right] \ , 
\end{eqnarray}
where the factor $s=2$ accounts for spin degeneracy. In 
what follows it is useful to use the approach of Altland and Gefen [18]
and to write
\begin{eqnarray}
G_{R,A} (\epsilon, B) &=& \frac{\partial}{\partial \epsilon}
\ln (\epsilon \pm i \eta - H_0 -V)=\nonumber \\
&=& \frac{\partial}{\partial \epsilon} \left[ \ln (\epsilon \pm i \eta - 
H_0)
+ \sum^\infty_{n=1} \frac{1}{n} (G^{(0)}_{R,A} V)^n\right] \ , 
\end{eqnarray}
where $H_0 = (1/2m)(p-\frac{e}{c}A)^2$ is the unperturbed 
Hamiltonian, $G^{(0)}_{R,A}$ are the corresponding Green's functions and
$V$ is the random potential. Eq.(9) is written in an operator form,
rather than in the coordinate representation. Averaging of a product
$G_R(\epsilon_1,B_1)G_A(\epsilon_2,B_2)$ results in the usual ``diffuson'' and
``cooperon'' series. The contribution of the $n$-th term of the cooperon
(diffuson) series to the correlation function $K(\epsilon_1,\epsilon_2;
B_1,B_2)$ is:
\begin{eqnarray}
K^{(n)}_\pm = \frac{s^2}{2\pi^2 V^2_0}
\frac{\partial^2}{\partial \epsilon_1\partial \epsilon_2}
\left[\frac{1}{n} Re Tr P^n_\pm (\epsilon_1-\epsilon_2,B_\pm)\right] \ , 
\end{eqnarray}
where sign $+(-)$ corresponds to cooperon (diffuson), and $B_\pm = B_1\pm B_2$.
The operator $P_+$, in coordinate representation, is written as
\begin{eqnarray}
P_+ (\vec{r},\vec{r}';\epsilon_1-\epsilon_2,B_+) =
\frac{\hbar}{2\pi\nu_0\tau}
\langle G_R(\vec{r},\vec{r}';\epsilon_1,B_1)\rangle
\langle G_A(\vec{r},\vec{r}';\epsilon_2,B_2)\rangle \ , 
\end{eqnarray}
where $\nu_0$ is the free electron density of states (per spin), for
an infinite system at $B=0$. The expression for 
$P_-(\vec{r},\vec{r}';\epsilon_1-\epsilon_2,B_-)$ is 
obtained from Eq.~(11) by interchanging $\vec{r}$ and $\vec{r}'$, 
in the advanced Green's function. For classically weak magnetic fields,
i.e. when $\omega_c\tau \ll 1$,
\begin{eqnarray}
\langle G_{R,A}(\vec{r},\vec{r}';\epsilon,B)\rangle=
\langle G_{R,A}(\vec{r},\vec{r}';\epsilon,B=0)\rangle
\exp\left(\frac{ie}{\hbar}\int^{\vec{r}'}_{\vec{r}} 
\vec{A}\cdot d\vec{\ell}\right) , 
\end{eqnarray}
where the integration is along a straight line connecting $\vec{r}$
to $\vec{r}'$.

To obtain the full function $K(\epsilon_1,\epsilon_2;B_1,B_2)$
one has to sum $K^{(n)}_\pm$ over $n$. Since, eventually, derivatives
with respect  to $B_1,B_2$ are to be taken [Eq.~(7)], one is 
interested only in the $B$-dependent part of $K$. The (single) term
with $n=1$ and the two terms with $n=2$ (one for diffuson, one for
cooperon) do not depend on $B$. This happens because the $B$-dependent
phase factor drops out in a diagonal term $\langle G_R(\vec{r},\vec{r})\rangle$,
as well as in the product 
$\langle G_R(\vec{r},\vec{r}')\rangle\langle G_R(\vec{r}',\vec{r})\rangle$
(and similarly for the advanced Green's functions). Thus, the 
final expression for the $B$-dependent part, $\Delta K$, of $K$ is:
\begin{eqnarray}
&&\Delta K (\epsilon_1,\epsilon_2; B_1,B_2)=
\frac{s^2}{2\pi^2 V_0^2}\frac{\partial^2}{\partial\epsilon_1\partial\epsilon_2}
Re \sum^\infty_{n=3} \frac{1}{n}\sum_\alpha
\left(\lambda^n_\alpha(\epsilon_1-\epsilon_2,B_+) + \right. \nonumber \\
&&\hspace*{2cm}\left. + \lambda^n_\alpha(\epsilon_1-\epsilon_2,B_-)\right)
\ , 
\end{eqnarray}
where $\lambda_\alpha(\epsilon_1-\epsilon_2,B_\pm)$ is the
$\alpha$-th eigenvalue of the operators $P_\pm$.

Eq.~(13), supplemented by the definition (11) of the operators $P_\pm$, forms
the basis for a quantitative treatment of orbital magnetism in
disordered metals, in a broad range of temperatures and fields.
The treatment is not restricted [18] to the standard diffusion
approximation for cooperon and diffuson. The situation is the same as in the
treatment of the weak localization correction for the conductivity.
This correction is expressed in terms of the eigenvalues of the operator
$P_+$ and, in order to obtain reasonably accurate quantitative results
in a broad range of magnetic fields, one has to evaluate these eigenvalues beyond
the diffusion approximation [15, 19-21].

Physical quantities of interest can be expressed in terms of 
$\Delta K$, given in Eq.~(13). Along with the variance
\begin{eqnarray}
\langle \Delta \chi^2\rangle =
\lim_{\limes}
\frac{\partial^4}{\partial B_1^2\partial B^2_2} \int\!\!\int d \epsilon_1
d\epsilon_2 F_T(\epsilon_1-\mu) F_T(\epsilon_2-\mu)
\Delta K(\epsilon_1,\epsilon_2;B_1,B_2) \ , 
\end{eqnarray}
one can consider various correlation functions, such as
$\langle \Delta \chi(B)\Delta\chi(B+\Delta B)\rangle$ or
$\langle \Delta \chi(T)\Delta\chi(T+\Delta T)\rangle$.
The first function is defined by Eq.~(14) with $B_1\rightarrow B$,
$B_2\rightarrow B+\Delta B$. To obtain the second function one
has to replace one of the $F_T$-factors in Eq.~(14) by $F_{T+\Delta T}$.
Other correlation functions, for susceptibility or magnetization, 
can be defined in a similar way.

Integration in Eq.~(14) can be carried out. After integration by parts, 
the derivatives contained in $\Delta K$ act on the $F_T$-factors,
giving
\begin{eqnarray}
\frac{d}{d\epsilon}F_T(\epsilon-\mu)=
\left[ \exp(\frac{\epsilon-\mu}{T}) + 1 \right]^{-1}\equiv
f_T(\epsilon-\mu) \ . 
\end{eqnarray}
The subsequent integration reduces to the Matsubara sums over the poles of
$f_T(\epsilon-\mu)$. This introduces a factor $(2\pi T)^2$
and changes the argument $\epsilon_1-\epsilon_2$ of the functions
under the integral to $\epsilon_p + \epsilon_q$ where $\epsilon_p=i \pi T
(2p + 1)$, $\epsilon_q=i\pi T(2q + 1)$ and $p,q=0,1,\dots$. The final
result is (the double sum over fermionic frequencies can be replaced
by a single sum over a bosonic frequency):
\begin{eqnarray}
&&\langle \Delta \chi^2\rangle =
\frac{2 s^2T^2}{V_o^2} \lim_{\limes}
\frac{\partial^4}{\partial B_1^2\partial B^2_2} Re \sum^\infty_{n=3}
\frac{1}{n} \sum^\infty_{p,q=0} \sum_\alpha
\left[\lambda^n_\alpha (\epsilon_p+\epsilon_q,B_+) + \right. \nonumber \\
&&\left. \hspace*{2cm} + \lambda^n_\alpha(\epsilon_p+\epsilon_q,B_-)\right] 
\end{eqnarray}
In the next Section we study $\langle \Delta\chi^2\rangle$, for various temperatures
and fields, and briefly discuss some correlation functions.\\

\noindent{\bf 5. Results and Discussion}

One should distinguish between weak magnetic fields, $L_B > L_T$ (linear response)
and stronger fields, when the dependence of magnetization on $B$ becomes
non-linear. Note that $L_B> L_T$ is a sufficient condition for linear response,
i.e. there is no requirement that $L_B$ should be larger than the sample size
$L$ (recall that we are interested in the ``high temperature'' case, 
$L_T < L$).  In this respect the situation is like in the thermodynamic
perturbation theory [1], where finite temperature stabilizes the perturbative
treatment and enlarges its range of validity. Thus, while computing the eigenvalues
of the operators $P_\pm$, we can assume $L_B<L$. Moreover, it is sufficient to
find only the eigenvalues of $P_+$. This is because the function $P_-$ differs
from $P_+$ only by its argument, containing $(B_1-B_2)$ instead of 
$(B_1+B_2)$. In particular, since in the definition of $\langle \Delta \chi^2\rangle$
the difference $(B_1-B_2)$ is  infinitesimal, the contribution of 
$P_-$ to $\langle \Delta \chi^2\rangle$ is the same as of $P_+$ in the limit
of small $B$.

Eigenvalues of the operator $P_+$ are well known, in connection with the weak
localization problem [20] (a closely related operator appears also in the theory of
superconductivity [22,23]). For the two-dimensional case, which is under
consideration here, the $N$-th eigenvalue of $P_+(\epsilon_1-\epsilon_2,B_+)$
is
\begin{eqnarray}
\Pi_N(\epsilon_1-\epsilon_2,B_+) =
\frac{L_+}{\ell} \int^\infty_0 dx L_N(x^2)
\exp\left[ - \frac{x^2}{2} - \frac{L_+}{\ell}x
(1-i\frac{\epsilon_1-\epsilon_2}{\hbar}\tau)\right] 
\end{eqnarray}
and each eigenvalue is degenerate $(L^2/\pi L^2_+)$ times. $L_N$ is the
Laguerre polynomial ($N=0,1,\dots$) and $L^2_+ = 2\hbar c/eB_+$.
Thus, the contribution of $P_+$ to the variance of susceptibility, 
Eq.~(16), is: \begin{eqnarray}
\langle \Delta \chi^2\rangle_+ =
\frac{2 s^2T^2}{L^4} \lim_{\limes}
\frac{\partial^4}{\partial^2 B_1\partial^2 B_2} 
\left[\frac{L^2}{\pi L^2_+}
Re \sum^\infty_{p,q=0}
\sum^\infty_{n=3} \sum^\infty_{N=0}
\frac{1}{n} \Pi^n_N(\epsilon_p+\epsilon_q,B_+)  \right] 
\end{eqnarray}
where
\begin{eqnarray}
\Pi_N(\epsilon_p + \epsilon_q, B_+) = 
\frac{L_+}{\ell} \int^\infty_0 dx L_N (x^2)\exp\left\{
-\frac{x^2}{2} - \frac{L_+}{\ell} x \left[ 1 + \frac{2\pi T\tau}{\hbar}
(p+q+1)\right]\right\} \ . 
\end{eqnarray}
To obtain the contribution $\langle \Delta \chi^2\rangle_-$, 
corresponding to $P_-$, one has to replace $B_+$ by $B_-$, in the argument
of $\Pi_N$ and in the definition of $L_+$. Thus, for any value of the actual
magnetic field $B$ (satisfying, of course, $\omega_c\tau \ll 1$), 
$\langle \Delta \chi^2\rangle_-$ is given by the weak field limit of
$\langle \Delta \chi^2\rangle_+$.

For $L_B,L_T \gg \ell$ (the diffusion regime), the integral in (18) is dominated
by small $x$, so that $L_N(x^2) \simeq 1-Nx^2$, and \
\begin{eqnarray}
\Pi^{\rm(dif)}_N \approx 1 - \frac{2\ell^2}{L^2_+} (N+\frac{1}{2}) -
\frac{2\pi T\tau}{\hbar} (p+q+1) \ , 
\end{eqnarray}
which holds for $N\ll (L_+/\ell)^2$ and $p,q \ll \hbar/T \tau$. Under these conditions
the sum over $n$ in Eq.~(18) can be approximated by $-\ln (1-\Pi_N)$ and
the expression for the typical value
$\langle \Delta \chi^2\rangle^{\frac{1}{2}}\equiv \delta \chi$ can be reduced to
the one given in [5]. In the limit of weak field
\begin{eqnarray}
\delta \chi = 1.67 \chi_0 \frac{L_T}{L} k_F\ell \ , 
\end{eqnarray}
where $\chi_0 = e^2/12 \pi mc^2$.

It is known from the previous studies [15,18-21] that calculations based
on the diffusion approximation for $\Pi_N$ are quantitatively accurate
only when the ratio $\ell/L_B$ is extremely small. In order to study  larger
fields or higher temperatures, in particular when $L_B$ or $L_T$ become smaller
than $\ell$ (the ballistic  regime), one must treat eigenvalues $\Pi_N$ more
accurately. Without resorting to numerics, one can use the approximate
expression [20]
\begin{eqnarray}
\Pi_N = \left[ 4\frac{\ell^2}{L^2_+} (N+\frac{1}{2}) + (1+\frac{2\pi T\tau}{\hbar}
(p+q+1))^2\right]^{-\frac{1}{2}} \ ,  
\end{eqnarray}
which interpolates between the two regimes. In the ballistic regime, as opposed
to the case of diffusion, $\Pi_N$ is small so that it is sufficient to keep 
in Eq.~(18) only the term $n=3$. Note that the present approach, while
giving the correct asymptotic dependence of $\delta \chi$ on $B$ and $T$, is not
capable of producing the correct numerical coefficient in the ballistic regime.
The reason is the that the ballistic regime is dominated by three-impurity diagrams,
and there are several such diagrams besides the $n=3$ diagram of the diffuson
or cooperon series.

Eq.~(18), together with the approximate expression (22) for $\Pi_N$, enables one to
consider a broad range of temperatures and fields. Let us first discuss the
case of linear response, i.e. $L_B>L_T$. In this case, as explained above,
$\langle \Delta \chi^2\rangle_+ = \langle \Delta \chi^2\rangle_-$ and does
not depend on $B$. At low temperatures, $L_T \gg \ell$, one is back to the 
diffusion regime, Eq.~(21).  At high temperatures $(L_T<\ell$, i.e. $T>\hbar/\tau)$
one enters the ballistic regime, where $\Pi_N$ is small.  Keeping only the
term $n=3$ in Eq.~(18), one has to compute the sum
\begin{eqnarray}
\sum^\infty_{N=0}\left[ 4b(N+\frac{1}{2}) + t^2\right]^{-\frac{3}{2}} 
\end{eqnarray}
where $b\equiv (\ell/L_+)^2$ and $t\equiv (2\pi T\tau/\hbar)(p+q+1)$.
Although the sum can be computed by the Euler-McLaurin formula, it is 
instructive to extract the  $T$-dependence of 
$\langle \Delta \chi^2\rangle$ by simple power counting. Since there are four
derivatives with respect to $B_1,B_2$ in Eq.~(18), one has to look for terms proportional
to $B^4_+$. One power of $B_+$ comes from the term $L^{-2}_+$, and the other
three from the function defined by the sum in Eq.~(23). Since this function
is of the form $t^{-3} f(b/t^2)$, it is clear that the term proportional to
$b^3$ will be multiplied by $t^{-9}$. The  sum over $p$ and $q$ rapidly converges, so
that
\begin{eqnarray}
\langle \Delta \chi^2\rangle  \simeq \frac{T^2}{L^2}
\frac{\partial^4}{\partial B^4} \left[\left(\frac{2  \pi T \tau}{\hbar}\right)^{-9}
\ell^6\left(\frac{eB}{\hbar c}\right)^4\right] \ , 
\end{eqnarray}
i.e. $\delta \chi \simeq \chi_0 k_F\ell(\ell/L)(\ell_T/\ell)^{\frac{7}{2}}$,
in agreement with the arguments of Sec.~3.

Consider now the opposite case, $L_B < L_T$, when 
$\langle \Delta \chi^2\rangle_+$ (but not
$\langle \Delta \chi^2\rangle_-$) begins to depend on $B$. Since
$\langle \Delta \chi^2\rangle_+$ decreases with $B$, it follows that for 
sufficiently strong fields
$\langle \Delta \chi^2\rangle$ will reach half of its low-field value. In 
the diffusion regime, 
$\ell << L_B << L_T$,
$\langle \Delta \chi^2\rangle_+$ decays rather slowly,  on a typical scale
of $\Delta B \simeq \phi_0/L^2_T$. For stronger fields,
$L_B << \ell << L_T$, the ballistic regime is reached. In this regime
$\langle \Delta \chi^2\rangle_+$ becomes negligible, compared to
$\langle \Delta \chi^2\rangle_-$, and keeps decreasing as $L^7_B$. Indeed,
in this case the sum (23) is of order $b^{-3/2}$, as long as 
$p,q < \sqrt{b} (\hbar/2\pi T \tau)$, and the expression in the square brackets
in Eq.~(18) is estimated as 
$(L^2/L^2_+)b^{-3/2} b(\hbar/2\pi T\tau)^2 \sim \sqrt{B_+}$. 
After differentiating four times with respect to $B_+$, one obtains
$\langle \Delta \chi^2\rangle_+ \simeq \chi^2_0 (k_F\ell)^2
(L^7_B/L^2\ell^5)$. Thus, just on the border of the ballistic regime
$(L_B\simeq \ell)$,
$\langle \Delta \chi^2\rangle_+ \simeq \chi^2_0(k_F\ell)^2(\ell/L)^2$,
i.e. is reduced by a factor $(\ell/L_T)^2$ as compared to its low-field value.

Various correlation functions can be studied in a similar way. Consider
the functions
$\langle(\Delta\chi(B)\Delta\chi(B+\Delta B)\rangle\equiv C(B,\Delta B)$.
This function, as was mentioned in the previous Section, differs from
$\langle \Delta \chi^2\rangle$ only by the limiting value of $B_2$, which is
now equal to $B+\Delta B$. Therefore, the ``cooperon part'', $C_+$,
is a function of $2B+\Delta B$, whereas the ``diffuson part'', $C_-$, is a function
of $\Delta B$. Both $C_+$ and $C_-$ decay with $\Delta B$. In addition,
$C_+$ decays as a function of $B$ and becomes negligible for $L_B\ll L_T$.
Thus, for a fixed temperature $(L_T \gg\ell)$ there should exist aperiodic oscillations
of magnetic susceptibility, in an individual (typical) sample, with a 
characteristic period $\Delta B \simeq \phi_0/L^2_T$ and amplitude
$\delta \chi \simeq \chi_0 k_F\ell(L_T/L)$.
Since $C_-(\Delta B)$ does not depend on $B$, these oscillations will
persist into the ballistic regime, $L_B \ll \ell$, and are limited only
by the condition $\omega_c \tau \ll 1$. The possible existence of such oscillations,
in the diffusion regime, was already mentioned in [6], although it was not realized
that these oscillations do not decay even in the ballistic regime.
Similar oscillations should exist, when the temperature or the
chemical potential are changed by an amount of order $T$ (or $E_c$, if
$T < E_c)$. Such a  change produces a ``new sample'', and, thus, a 
corresponding change in the susceptibility.\\

\noindent{\bf 6. Interactions}

Interactions can have a profound effect on orbital magnetism. Long ago
Aslamazov and Larkin [24] pointed out the existence of an interaction 
induced paramagnetism. This phenomenon is similar to the interaction induced
diamagnetism in superconductors above the critical temperature. Since,
however, in a normal metal the interaction is repulsive, one ends up with a 
paramagnetic term. In two dimensions this term supercedes Landau diamagnetism of non-interacting
electrons.

Static disorder modifies but does not destroy the effect. The combined
effect of interactions and disorder was studied in detail in [25,26], where the
average
susceptibility for an ensemble of disordered samples was  calculated
(for a semiclassical derivation see [27]). The
effect is due to the interaction (in the presence of disorder) between a pair
of electrons in the Cooper channel. This interaction, at zero magnetic field,
produces a small correction to the single particle density of states. In two
dimensions the correction, at the Fermi energy $\mu$, is:
\begin{eqnarray}
\delta \nu_c \simeq \frac{1}{\hbar D} \ln \frac{\ln (T_0/T)}{\ln (T_0\tau/\hbar)}, 
\end{eqnarray}
where $T_0 = \mu\exp(1/\lambda_0)$ and $\lambda_0$ is the effective
electron-electron interaction constant. If $\lambda_0$ were much smaller
than $1/\ln(\mu/T)$, one could expand the ratio of the logarithms in
$\lambda_0$, to obtain the first order result $\delta \nu_c\simeq (\lambda_0/\hbar D)
\ln (T\tau/\hbar)$. Since, however, the aforementioned condition is violated,
one should use the renormalized interaction in the Cooper channel which results
in the double logarithm in Eq.~(25), with $T_0$ roughly equal to $\mu$.

The magnitude of the interaction induced paramagnetic susceptibility (in the
linear response regime) can be estimated by using the general relation
between susceptibility $\chi$ and the value of the effective Bohr magneton
$\beta_{\rm eff}$ for particles in question, namely: $\chi \simeq 
\beta^2_{\rm eff}\nu$, where $\nu$ is the corresponding density of states.
The ``particles'' (cooperons) have charge  2e and ``mass'' $\hbar/2D$, so that
$\beta_{\rm eff} \simeq e D/c$ (larger by a factor $k_F\ell$ than the electron
Bohr magneton $e\hbar/2mc)$. The corresponding density of states is the
correction $\delta \nu_c$, Eq.~(25). Thus, the average paramagnetic
susceptibility is
\begin{eqnarray}
\langle \chi_p\rangle \simeq \chi_0 k_F\ell \ln\frac{\ln(\mu/T)}{\ln(\mu\tau/\hbar)}
 \ . 
\end{eqnarray}

Note that a similar estimate could have been made for the typical
susceptibility of a mesoscopic sample, in the absence of interaction
(Sec.~3). Indeed, multiplying $\beta^2_{\rm eff}$ by the mesoscopic correction
to the density of states, $|\delta \nu|\simeq 1/\hbar D$, gives $\chi \simeq k_F\ell
\chi_0$. Let us emphasize that there is a big difference between the mesoscopic
correction $\delta \nu$ and the interaction induced correction $\delta \nu_c$. The
former is a finite size effect, sensitive to temperature and rapidly
oscillating with $\mu$. The latter is a robust, intrinsic property of
the interacting system, so that the interaction induced paramagnetism
 is not at all a mesoscopic effect.

The large paramagnetic susceptibility remains, of course, present also in the
mesoscopic samples considered in Sec.~3-5. For $T>E_c$ the average
susceptibility $\langle \chi \rangle$ is already close to its
macroscopic limit, Eq.~(26). (For $T \ll E_c$ some finite size effects in
$\langle \chi \rangle$ show up [4].) Thus, the mesoscopic fluctuations
discussed in the present paper will occur on the background of $\langle 
\chi_p\rangle$. These fluctuations do not seem to be affected by interactions
in a significant way,  similarly to the case of the universal conductance
fluctuations, as discussed in [16].  However, the full picture
of the mesoscopic magnetism in disordered metals, in the presence of interactions
and for arbitrary temperatures and fields, 
has not yet emerged. \\

\noindent{\bf 7. Conclusions}

The theory of mesoscopic effects in orbital magnetism of disordered metals
was reviewed and extended to higher temperatures and fields. The effects
decay with temperature rather gradually and persist to temperatures
much higher than the Thouless energy $E_c$. One manifestation of mesoscopic
orbital magnetism is aperiodic, sample-specific oscillations in magnetization
and susceptibility under changes of various external factors -- temperature,
chemical potential or external magnetic field. The oscillations persist
to quite strong fields, limited only by the condition $\omega_c\tau \ll 1$.
The amplitude of oscillations in the susceptibility is of order $\chi_0 k_F\ell$
at $T\simeq E_c$, and slowly decreases at higher
temperatures.

One should distinguish between the average susceptibility, measured
on a large ensemble of macroscopically identical samples, and the typical
susceptibility measured on an individual sample. I am not aware of any 
measurements of orbital magnetism in disordered mesoscopic metals. The similar
phenomenon of orbital magnetic response to an Aharonov-Bohm flux has been 
investigated, to some extent, experimentally [28-31]. In particular,
measurements on single rings, disordered [29] and quasiballistic [30], were made.
Since, at low temperatures and weak fields, the magnetic moment of a ring is about
the same as of a singly-connected sample of a comparable size, the mesoscopic
fluctuations studied in the present paper should be amenable to observation.

Some aspects of the phenomenon of mesoscopic orbital magnetism were not
covered in this work. We have not discussed mesoscopic effects for
fields $\omega_c\tau > 1$, i.e. in the regime of the de-Haas-van Alphen oscillations
or in the quantum Hall regime. Another omission is the mesoscopic magnetism in three
dimensions, where a qualitatively new effect shows up. The susceptibility
of an individual sample becomes a tensor $\chi_{\alpha\beta}$ $(\alpha,\beta=x,y,z)$,
with non-zero off diagonal components, as opposed to the ensemble averaged
susceptibility. This happens because only after averaging over the impurities
the sample becomes spatially homogeneous. In an individual sample the
direction of magnetization does not coincide with the direction of the
magnetic field. For instance, for a sample of a spherical shape, the ratio
between 
$\langle \Delta \chi^2_{xz}\rangle^{\frac{1}{2}}$ and 
$\langle \Delta \chi^2_{zz}\rangle^{\frac{1}{2}}$, is $1/\sqrt{3}$
[5,6]. It would be of interest to investigate different sample shapes and
to extend the treatment beyond linear response. \\

\noindent{\bf Acknowledgements}

Part of this work was done during the "Extended Research Workshop on
Disorder, Chaos and Interaction in Mesoscopic Systems", in Trieste. I
am  grateful to the organizers for hospitality and to some of the
participants - I. Aleiner, B. Altshuler, V. Ambegaokar, Y. Gefen, A. Larkin, 
P. Mohanti, A. Zyuzin - for useful discussions and explanations.
Illuminating conversations with F. von Oppen, L. Pitaevskii and
U. Sivan are gratefully  acknowledged. The research was supported
by the Fund for Promotion of Research at the Technion. \\

\noindent{\bf References}
\begin{enumerate}
\item Landau L.D. and Lifshitz E.M. 1980 Statistical Physics, 3rd Edition,
Part 1 (Pergamon Press).
\item Shapiro B. 1993 Physica {\bf A200}, 498 (Proc.\ 4-th International 
Conference on Frontiers in Condensed Matter Physics); Semiconductors {\bf 27},
467 (Proc. IPCMP '92).
\item Fukuyama H. 1989 Jour.\ Phys.\ Soc.\ Japan {\bf 58}, 47.
\item Oh S., Zyuzin A. Yu and Serota R.A. 1991 Phys.\ Rev.\ {\bf B44}, 8858.
\item Raveh A. and Shapiro B. 1992 Europhys.\ Lett.\ {\bf 19}, 109 (Erratum p. 565).
\item Raveh A. 1993 Orbital Magnetic Response of Disordered Conductors, Ph.D. 
Thesis (Technion, Haifa).
\item von Oppen F. and Riedel E.K. 1993 Phys.\ Rev.\ {\bf 48}, 9170.
\item Argaman N., Imry Y., and Smilansky U. 1993 Phys.\ Rev.\ {\bf B47}, 4440.
\item Gutzwiller M.C. 1990 Chaos in Classical and Quantum Mechanics (Springer,
Berlin).
\item Agam O. 1994 J.\ Phys.\ I France {\bf 4}, 697.
\item Bogachek E.N. and Gogadze G.A. 1972 Sov.\ Phys.\ JETP {\bf 36}, 973.
\item Richter K., Ullmo D., and Jalabert R.A., 1996 Phys.\ Rep.\ {\bf 276}, 1.
\item von Oppen F. 1994 Phys.\ Rev.\ {\bf B50}, 17151.
\item Altshuler B.L., Gefen Y., Imry Y., and Montambaux G. 1993 Phys.\ Rev.\
{B47}, 10335.
\item Dyakonov M.I. 1994 Sol.\ St.\ Communications {\bf 92}, 711.
\item Lee P.A., Stone A.D., and Fukuyama H. 1987 Phys.\ Rev.\ {\bf B35},
1039.
\item Altshuler B.L. and Shklovskii B.I. 1986 Sov.\ Phys.\ JETP {\bf 64}, 127.
\item Altland A. and Gefen Y. 1995 Phys.\ Rev.\ {\bf B51}, 10671.
\item Gasparian V.M. and Zyuzin A.Yu. 1985 Sov.\ Phys.\ Solid State {\bf 27},
999.
\item Kawabata A. 1984 J.\ Phys.\ Soc.\ Japan {\bf 53}, 3540.
\item Cassam-Chenai A. and Shapiro B. 1994 J.\ Phys.\ I France {\bf 4}, 1527.
\item Lee P.A. and Payne M.G. 1972 Phys.\ Rev.\ {\bf B5}, 923.
\item Kurkijarvi J., Ambegaokar V. and Eilenberger G. 1972 Phys.\ Rev.\ {\bf B5},
868.
\item Aslamazov L.G. and Larkin A.I. 1974 Sov.\ Phys.\ JETP {\bf 40}, 321.
\item Altshuler B.L., Aronov A.G. and Zyuzin A.Yu. 1983 Sov.\ Phys.\ 
JETP {\bf 57}, 889.
\item Altshuler B.L. and Aronov A.G. 1985 in Electron-Electron Interactions in 
Disordered Systems, ed. Efros A.L. and Pollak M. (North-Holland).
\item Ullmo D., Richter K., Baranger H.U., von Oppen F., and Jalabert R.A.
1997 Physica {\bf E1}, 268.
\item Levy L.P., Dolan G., Dunsmuir J., and Bouchiat H. 1990 Phys.\ Rev.\
Lett. {\bf 64}, 2074.
\item Chandrasekhar V., Webb R.A., Brady M.J., Ketchen M.B., Gallagher W.J., and
Kleinsasser A. 1991, Phys.\ Rev.\ Lett.\ {\bf 67}, 3578 (1991).
\item Mailly D., Chapelier C., and Benoit A., 1993 Phys.\ Rev.\ Lett.\ {\bf 70},
2020.
\item Mohanty P., Jariwala E.M.Q., Ketchen M.B. and Webb R.A., preprint
\end{enumerate}

\end{document}